# WeDo:
# Exploring Participatory, End-To-End Collective Action

Haoqi Zhang, Aaron Shaw, and Liz Gerber, Northwestern University
Andrés Monroy-Hernández, and Shelly Farnham, Microsoft Research
Sean Munson and Benjamin Mako Hill, University of Washington
Peter Kinnaird, Carnegie Mellon University
Patrick Minder, University of Zurich

## 1. INTRODUCTION

Collective intelligence and social computing systems play an increasingly important role in collective action—that is, "actions taken by two or more people in pursuit of the same collective good" [Marwell and Oliver 1993]. Social computing platforms that advance collective action connect crowds and communities of participants, lower the cost of communication, facilitate deliberation, and help to coordinate action, thereby enabling new forms of collaboration and collective mobilization.

Numerous social computing platforms facilitate distinct pieces of collective action or context-specific collaborations. We take inspiration from these platforms to design a system, *WeDo*, aimed at supporting simple *participatory* and *end-to-end* collective action, in which a crowd or community comes together to surface opportunities, formulate goals, brainstorm ideas, make plans, and mobilize a critical mass of participants to take action.

We created and deployed WeDo as a technology probe [Hutchinson et al. 2003] to identify possibilities for, and barriers to, building and deploying participatory, end-to-end collective action systems. A Twitter app combined with a complementary web interface (see Figure 1), WeDo supports: the creation of high-level "missions;" collecting ideas from a community for accomplishing the mission; browsing and voting on ideas; and following through on actions. The system automates the transition through these stages, so as to promote progress and follow-through. Our experience with WeDo underscores opportunities and challenges for designers of end-to-end CSCA systems.

## 2. RELATED WORK

Existing systems supporting collective action tend to fall into one or more of the following categories:

**Systems Supporting Few Tasks.** Many systems focus exclusively on a few, specific tasks within larger processes. For example, *FixMyStreet* allows people to submit reports of issues such as potholes, broken lights, or graffiti [King and Brown 2007].

**Systems Supporting Few Stages.** Systems such as *Stanford Catalyst* help community members share activities and sign up for activities they wish to participate in, but do not allow users to plan the details of another's suggested activity nor to brainstorm about possible activities collaboratively.

**Systems Supporting Action in Few Contexts.** Another class of systems rely on a narrowly defined context or domain to structure group activities. For example, *GitHub* combines code repositories, bug trackers, and discussion boards to allow software developers to surface issues, deliberate, submit resolutions, and adopt them [Schweik and English 2012], activities defined by the domain that rarely translate well to other contexts.

**Systems Supporting Less Participatory Actions.** Systems such as *Kickstarter* help people market projects and recruit supporters, who provide funds but do not develop the projects directly.

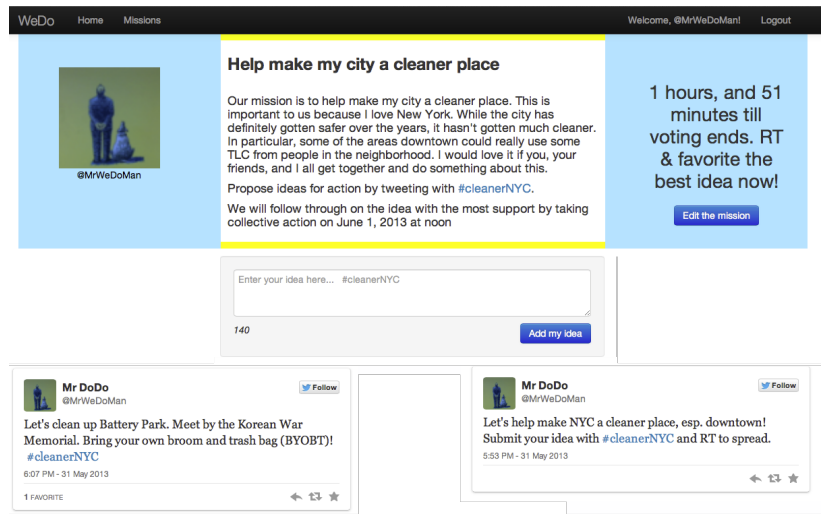

Fig. 1. Web interface for viewing and contributing to missions in WeDo. Users can also contribute directly via Twitter.

These systems raise the question of whether it is possible to create systems that support participatory, end-to-end collective action without relying on features designed for domain-specific constraints. Such systems promise potential enhancements over existing tools by (1) distributing tasks more effectively throughout the process of collective action through collective intelligence and cooperation [Benkler 2006; Woolley et al. 2010]; (2) facilitating the progression toward eventual action by managing transitions across discrete types of tasks [González and Mark 2004]; and (3) meeting needs of communities of end users who may benefit from participatory, end-to-end processes of collective action.

## 3. THE WEDO PROTOTYPE

*WeDo* is a lightweight system for promoting participatory, end-to-end collective action. WeDo supports collective action in four phases: start a "mission," collect ideas, vote on a plan for action, and notify people to coordinate action. The prototype consists of a web interface for submitting, tracking, and participating in missions, as well as a Twitter bot that automates the distribution of messages for ideation, voting, selection, and action notification.

For example, a participant could create a mission to cleanup a local park. WeDo announces the mission via posts to the participant's Twitter account as well as the WeDo account, soliciting suggestions. After a few hours, the system prompts interested participants to vote on promising ideas by retweeting or favoriting those they like. After the voting deadline, the system sends messages announcing the leading choice and prompting the submission of additional details. Finally, the system notifies participants immediately before the event that it is about to occur.

Building on top of Twitter has many advantages. First, within any phase of the process, WeDo allows anyone to discuss ideas and refine plans directly on Twitter. Second, as a messaging system, Twitter makes it easy for people to contribute and to receive notifications anywhere, which facilitates transitioning through different phases of collective action. Third, functionalities for expressing support or voting are already built in via retweeting and favoriting. Finally, the social network in Twitter can kickstart missions by leveraging existing followers and conversations.

To supplement Twitter's affordances, the WeDo prototype also includes a companion web interface for visualizing ideas and coordinating actions. The interface provides more context for a mission than Twitter and supports more structured interactions. In the current version, a participant creates a mission by filling out a form with fields for the name of the mission, a description prompted by the text, "this mission is important to me because," a hashtag for the mission, a date and time by which the idea is to be selected, and a date and time by which the mission is to be executed. When the form is submitted, the participant gets presented with a suggested tweet to start the mission. The message can be edited, but is constrained to the 140 character limit of Twitter.

Once the mission is created, interested participants can submit ideas either via Twitter or the WeDo web interface. Automated messages are sent out at their scheduled time. The web interface provides light support for browsing submitted ideas by placing the ideas with the most votes at the top. It also helps users track the process by grouping messages based on the stages, and showing the time left until the next stage (see Figure 1).

## 4. LESSONS LEARNED FROM INITIAL DEPLOYMENTS

We have conducted three initial deployments of WeDo to celebrate the end of a workshop, to do something together at a conference banquet, and to choose a book for a Twitter book club. Our deployment experiences (discussed in unpublished work currently under review) demonstrate several ways in which WeDo facilitates aspects of end-to-end collective action. First, many participants expressed that Twitter made adding ideas easy. They found voting on existing ideas through retweets and favorites intuitive, and liked being able to participate via mobile devices. Second, participants pointed out that the system helped to move the process along, steering them toward subsequent action. Third, participants appreciated how the system facilitates idea generation and decision-making. They understood that the decision was based on their votes, and that they generated the ideas that are voted on. Finally, even though most missions were open-ended and broad, a number of deployments were successful in that they resulted in participatory, end-to-end collective action processes.

The WeDo deployments also surfaced many challenges:

—*Fluidity and barriers between phases of collective action:* Fluidity between stages confused some participants, especially as the system relied on a single hashtag for many different actions.
—*Confounding user expectations and existing norms:* Participants expected other features from the system such as being able to vote on an idea suggested by multiple participants and having those votes merged (even when retweets were modified).
—*Clarity of high-level mission and task design:* Newcomers did not know how to contribute when mission goals were too broad or vague.
—*Identifying and mobilizing leadership:* Without explicit support for identifying and mobilizing leaders, missions can fail to gather the necessary momentum to move forward despite collective interest.
—*Opportunities for peripheral participation:* The expectation that participants will persist throughout all stages of a mission is prohibitive.
—*Platform trust and security concerns:* Requiring Twitter permissions may be prohibitive for participants unwilling or unable to share information.

## 5. CONCLUSION

Our experiences with the WeDo prototype recall the socio-technical gap [Ackerman 2000]. Existing tools serve as building blocks which may be assembled into more effective and comprehensive systems. In this way, the WeDo system highlights potential opportunities for social computing support collective action and provides insights to inform the design of future systems.